# VALIDATION OF THE 650 MHZ SRF TUNER ON THE LOW AND HIGH BETA CAVITIES FOR PIP-II AT 2 K*

C. Contreras-Martinez†, S. Chandrasekaran, S. Cheban, G. Eremeev, I. Gonin, T. Khabiboulline, Y. Pischalnikov, O. Prokofiev, A. Sukhanov, JC. Yun, Fermilab, Batavia, IL, USA

*Abstract*

The PIP-II linac will include thirty-six $\beta_G$=0.61 and twenty-four $\beta_G$=0.92 650 MHz 5 cell elliptical SRF cavities. Each cavity will be equipped with a tuning system consisting of a double lever slow tuner for coarse frequency tuning and a piezoelectric actuator for fine frequency tuning. The same tuner will be used for both the $\beta_G$=0.61 and $\beta_G$=0.92 cavities. Results of testing the cavity-tuner system for the $\beta_G$=0.61 will be presented for the first time.

## INTRODUCTION

The proton improvement plan (PIP)-II linac section under construction at Fermilab will consist of five classes of superconducting RF (SRF) cavities made of niobium. The linac will accelerate a proton beam to support experiments for the g-2, mu2e, and DUNE collaborations at Fermilab. Two types of elliptical cavities, the low beta (LB) cavity at $\beta_G$= 0.61 and the high beta (HB) cavity at $\beta_G$= 0.92, are used to accelerate the proton beam from 185 MeV to 800 MeV [1].

The SRF cavity tuner has three roles. It is needed for active microphonics compensation. It is also used for moving the cavities to the nominal frequency after cooling to 2 K. Lastly, it is used for protecting the cavity during pressure tests. This paper presents the double lever arm tuner testing on the cavity at 2 K. The cavity was placed in the recently upgraded cryostat [3] at the Meson Detector Building (MDB) at Fermilab, pictured in Fig. 1. This double lever tuner will be used for both the HB and LB 650 MHz elliptical cavities. The tuner specifications for the HB and LB 650 MHz cavities are shown in Table 1. There are two components to the tuner, one is the slow and coarse frequency tuning component consisting of a stepper motor. The other is the fast and fine frequency tuning component composed of piezoelectric actuators.

The double lever arm tuner's slow and coarse electromechanical component consists of a stepper motor manufactured by Phytron. Accelerated lifetime tests at Fermilab demonstrate that this stepper motor will survive prolonged operation far exceeding the typical linac lifetime of 25 years [4,5]. Additionally, after an irradiation hardness test (gamma rays), no performance degradation was observed, demonstrating that the stepper motor can survive under these operating conditions [5].

The fast and fine tuner component consists of two piezoelectric actuator capsules. The piezoelectric actuators are used for fast and fine frequency tuning control of microphonics. The piezoelectric actuators are designed and fabricated by Physik Instrumente (PI) per Fermilab specifications. The accelerated lifetime test demonstrates that the piezo can sustain $2 \times 10^{10}$ pulses with a peak-to-peak amplitude of 2 V. These are the number of cycles expected for the PIP-II linac, equivalent to 25 years. The irradiation test, with the same parameters as the stepper motor test, also showed minimal degradation for this actuator [5].

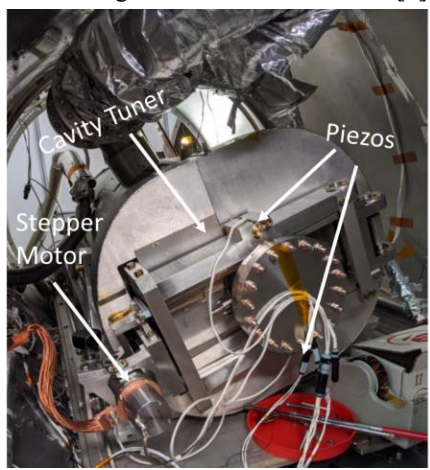

Figure 1: 650 MHz $\beta_G = 0.92$ with tuner and other ancillaries inside the STC cryostat at the MDB facility in Fermilab.

## SLOW AND COARSE TUNING COMPONENT

The tuner was tested on both the 650 MHz $\beta_G = 0.92$ and $\beta_G = 0.61$ dressed cavities. Once the cavity is cooled to 2 K the frequency will not be precisely at 650MHz. Since the tuner can only compress the cavity, which lowers the cavity frequency, the cavity frequency setpoint is set during the production to avoid being below 650 MHz at 2 K. The cavity's frequency at 2 K before tuning is shown in Table 2. This frequency is called the 2 K landing frequency ($f_{2\,K\,Landing}$), note that it is higher than 650 MHz. The slow coarse tuner has three regions of operation. The first region is when the tuner can stretch the cavity via the safety rods; in this region, the piezos are not engaged. The first region is not used during operation. The second region is the unrestrained position where the cavity frequency changes slightly or not at all. This region is due to the safety gap setup at room temperature. The piezos are not engaged in this region, hence the small frequency change. The last region is for normal operation where the piezos are engaged. These regions are shown in Fig. 2.

Table 1: 650 MHz cavity and tuner specifications for different geometries.

|  | $\beta_G = 0.92$ | $\beta_G = 0.61$ |
|---|---|---|
| Cavity Stiffness [kN/mm] | 5 | 4 |
| Cavity Tuning Sensitivity [Hz/$\mu m$] | 150 | 240 |
| Tuner System Stiffness [kN/m] | $\geq 40$ | $\geq 40$ |
| Lowest Mechanical Resonance of Cavity-tuner System [Hz] | >100 | >100 |
| Slow Tuner Frequency Range [kHz] | 200 | 200 |
| Stepper Motor Resolution [Hz/step] | ≤1 | ≤1 |
| Slow Tuner Hysteresis [Hz] | ≤ 100 | ≤ 100 |
| Piezo Tuner Frequency Range (at 120 V) [kHz] | 1.2 | 1.2 |
| Piezo Tuner Resolution [Hz] | <0.5 | <0.5 |

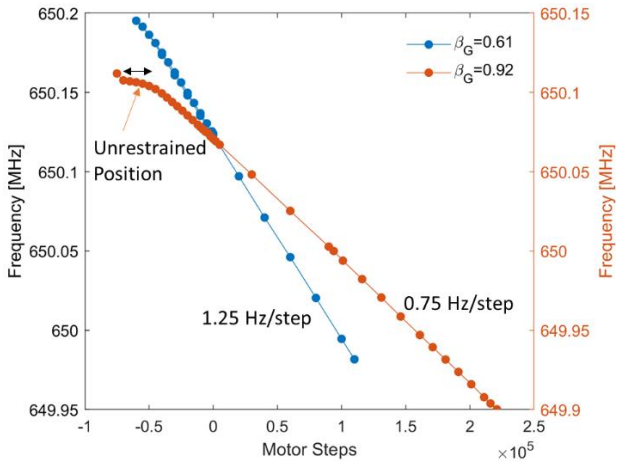

Figure 2: Tuner operation after cooldown to 2 K. The left axis corresponds to the $\beta_G$=0.61 and the right to $\beta_G = 0.92$.

The figures of merit of the 650 MHz tuner at 2 K are given in Table 2. The piezo preload frequency at 2 K is calculated from the difference between the 2 K landing frequency and the unrestrained frequency. The measured piezo preload at 2 K value is expected since the room temperature piezo preload was set to 30 kHz. The safety gap for the safety rod is shown in Fig. 2. This is the region where the frequency plateaus. The size of the frequency plateau gives the safety gap ( see Table 2). The safety gap value at room temperature was set to 150 μm. From the 2 K landing frequency, the number of steps needed to reach 650 MHz is given in Table 2. The total range of the tuner is also shown in Table 2, the value exceeds the specification given in Table 1. The $f_{unrestrained}$ value for $\beta_G = 0.92$ at room temperature was 648.959 MHZ. Note that the pressure in the helium vessel and outside the helium vessel was atm. The difference between the $f_{unrestrained}$ at 2 K and at room temperature is $1.148 \pm 0.1$ MHz.

The hysteresis of the stepper motor was tested by first operating it in short step increments and then in large step increments. The short-range hysteresis has increments of 10 steps, and the difference between the compression and relaxation sweep is 30 Hz, as shown in Fig. 3 for both cavity types. The 30 Hz value of the slow tuner hysteresis is consistent with the stepper motor actuator backlash measured with the LCLS-II tuner [6]. This is also within the hysteresis specification given in Table 1.

Table 2: Measured figures of merit of the 650 MHz cavity tuner at 2 K.

|  | $\beta_G = 0.92$ | $\beta_G = 0.61$ |
|---|---|---|
| f$_{2\ K\ Landing}$ [MHz] | 650.070 | 650.124 |
| f$_{unrestrained}$ [MHz] | 650.107 | 650.2 ± 20 kHz |
| Piezo preload 2 K [kHz] | 37 | TBD |
| Unrestrained gap 2K [μm] | 100 | TBD |
| Motor Steps to 650 MHz | 93333 | 99200 |
| Motor Sensitivity [Hz/Step] | 0.75 | 1.25 |
| Motor Range [kHz] | ≥212 | ≥214 |
| Piezo Sensitivity [Hz/V] | -24 | -36 |

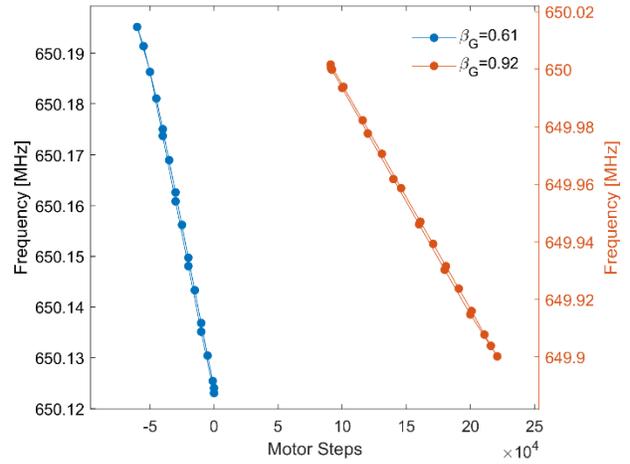

Figure 4: Large step stepper motor hysteresis.

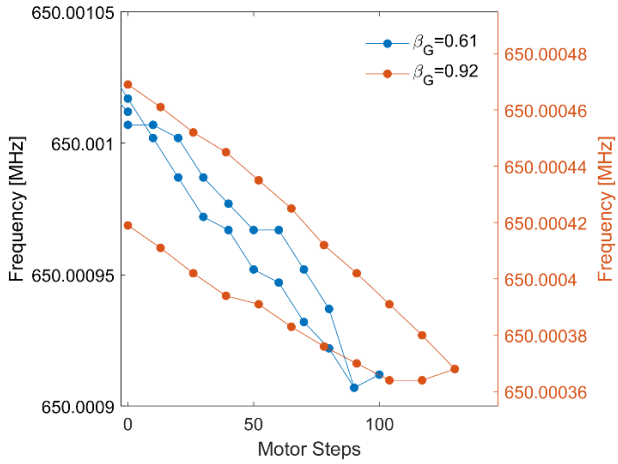

Figure 3: Short step hysteresis of the stepper motor for both cavity types.

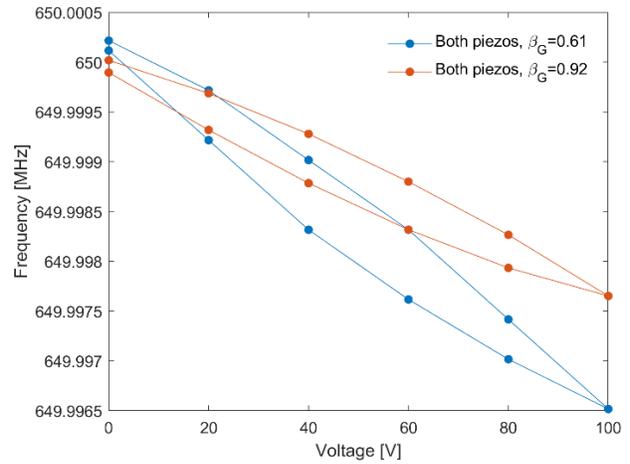

Figure 5: Piezo hysteresis for both types of cavities with 20 V intervals. Both piezos were used.

The tuner sensitivities in this range are given in Table 2 and are within the specification given in Table 1. The long-range hysteresis with a span of $10^8$ steps is shown in Fig. 4. These values demonstrate that the stepper motor can tune the cavity to 650 MHz and has a large range.

## FAST AND FINE-TUNING COMPONENT

The tuner consists of two piezo capsules which make contact with the cavity. The piezo actuator can expand by 34 ± 2 $\mu m$ when 100 V is applied at room temperature. The piezo displacement was studied on the 650 MHz cavity at 2 K, with results shown in Fig. 5 for both types of cavities.

The piezo can be modulated in small increments such as 15 mV achieving a piezo resolution of 0.5 Hz, which is within the specification in Table 1. At 100 V on both piezo capsules, the cavity frequency shift was -2.4 kHz for β=0.92 and -3.6 kHz for β=0.61 (see Fig. 5), which is higher than the specification given in Table 1. The expected detuning from microphonics of the cavity operated in CW mode is 5-50 Hz. Therefore, the piezo actuators provide a sufficient compensation range [8].

## CONCLUSION

The double lever tuner for the 650 MHz elliptical cavities ( β = 0.92 and β = 0.61 ) was tested at 2 K inside a cryostat on a dressed cavity. The results show that the slow-coarse range for the β = 0.92 cavity is larger than 212 kHz and larger than 214 kHz for the β = 0.61 cavity. The hysteresis for the slow tuner is 30 Hz, consistent with the results in [6] and within the specifications shown in Table 1. The fast-fine component test yielded a response of -24 Hz/V for β = 0.92 and -36 Hz/V for β = 0.61. This gives a large range for microphonics compensation. Thus, the double lever tuner is capable of operation in the PIP-II linac by exceeding the specification of the PIP-II project.

## ACKNOWLEDGMENTS

This manuscript has been authored by Fermi Research Alliance, LLC under Contract No. DE-AC02-07CH11359 with U.S. Department of Energy, Office of Science, Office of High Energy Physics.